\documentclass[rnote]{aa}
\usepackage{epsfig}
\usepackage{graphics}

\newcommand{\be}{\begin{equation}}
\newcommand{\ee}{\end{equation}}
\newcommand{\Msun}{M_\odot}

\begin{document}

\title{The Hubble diagram for a system within dark energy:\\ 
the location of the zero-gravity radius and the global Hubble rate}

\author{P. Teerikorpi\inst{1} \and A. D. Chernin\inst{1,2}}

\institute{Tuorla Observatory, Department of Physics and Astronomy, University of Turku, 21500 Piikki\"o,
Finland \and Sternberg Astronomical Institute, Moscow University, Moscow, 119899, Russia}

\authorrunning{P. Teerikorpi \& A.D. Chernin}
\titlerunning{Hubble diagram for a system within dark energy}

\date{Received / Accepted}

\abstract
{}
{Here we continue to
discuss the principle of the local measurement of dark energy using the normalized Hubble
 diagram describing
the environment of a system of galaxies.}
{We calculate the present locus of test particles injected a fixed time ago ($\sim$ the age of
the universe), in the standard
$\Lambda$ cosmology and for
different values of the system parameters (the model includes a central point mass $M$ and
a local dark energy density $\rho_{\rm loc}$) and
discuss the position of the zero-gravity distance $R_{\rm v}$ in the Hubble diagram.}
{Our main conclusion are: 1) When
the local DE density $\rho_{\rm loc}$ is equal to the global DE density $\rho_{\rm v}$,
the outflow reaches
the global Hubble rate at the distance $R_2 = (1+z_{\rm v})R_{\rm v}$,
where $z_{\rm v}$ is the global zero-acceleration redshift ($\approx 0.7$ for the
standard model).  This is also
the radius of the ideal Einstein-Straus vacuole.
2) For a wide range of the local-to-global dark energy ratio $\rho_{\rm loc}/\rho_{\rm v}$,
the local flow reaches the known global rate (the Hubble constant) at a distance $R_2 \ga 1.5 \times
R_{\rm v}$. Hence, $R_{\rm v}$  will be between $R_2/2$ and $R_2$, giving
upper and lower limits to $\rho_{\rm loc}/M$.
%in the Hubble diagram around a mass concentration when the Hubble constant is known.
For the Local Group, this supports the view that the local density is near the
global one.
%For the Local Group, $R_2 \approx 2$ Mpc, hence
%$(M_{\rm LG}/3.5\,10^{12}\Msun) < \rho_{\rm loc}/\rho_{\rm v} < 2.0
% \times (M_{\rm LG}/3.5\,10^{12}\Msun)$.
}
{}

\maketitle

\section{Introduction}

The standard $\Lambda$CDM
cosmology views dark energy as having
constant density everywhere. 
If so, its local density should be
identical to that inferred from global
observations of Supernovae Ia (Riess et al. \cite{riess98}, Perlmutter et al. \cite{perlmutter99})
and the CMB (Spergel et al. \cite{spergel07}),
$\rho_{\rm v}  \approx 7 \times 10^{-30}$ g/cm.

Dark energy can be studied locally (Chernin et al. \cite{chernin06}; Teerikorpi et al.
\cite{teerikorpi08}), as its
"antigravity" can affect galaxy motions near us
in a volume a few Mpc across (Chernin et al. \cite{chernin00}).
Also, the real
mass of a system can only be found if the dark energy is
included, because
gravitating systems
"lose" a part of their gravity due to the antigravity of the dark
energy within them (Chernin et al. \cite{chernin09}).

The homogeneity and constancy of the dark energy imply an important fact: lumps
 of matter, caused by gravitational instability in the expanding universe,
 did not appear alone, but together with zero-gravity surfaces around them. 
The gravity of the mass $M$ of the concentration and the repulsion of
 the dark energy with
density $\rho_{\rm v}$ are equal at the zero-gravity radius $R_{\rm v}$ (Chernin \cite{chernin01}):
\be
 R_{\rm v} = (\frac{3 M}{8\pi \rho_{\rm v}})^{1/3}. 
\ee
Antigravity dominates at $R > R_{\rm v}$, and gravity is stronger than
antigravity at $R < R_{\rm v}$.

This concept of the spatial border $R_{\rm v}$ does not apply to the perfectly
smooth Friedmann models
(these have instead the temporary border separating deceleration and acceleration everywhere), 
but it is important for understanding the local structure and dynamics of galaxy systems:
 the zero-gravity border exists during all the life-time of the system since its formation.

The two major parameters of a "local Hubble cell" are the
total mass of the group or cluster and the local density of the dark energy.
If the
mass is, say, $ 2 \times 10^{12} M_{\odot}$
and the local dark energy density is equal to its global value $\rho_{\rm v}$, then
$R_{\rm v} = 1.3$ Mpc.
% This shows that the group is located inside the
%sphere of radius $R_{\rm v}$ while the flow occurs outside the sphere. 
Or, if the zero-gravity radius is known from observations, the
mass may be written as:
% 8
\be M = {\frac{8\pi}{3}} \rho_{\rm v} R_{\rm v}^{3} \simeq 0.9 \times 10^{12}
[R_{\rm v}/({\rm Mpc})]^{3} M_{\odot}. \ee \noindent
Thus one can determine the DE density by comparing the predicted value of
the group mass to the mass known from other methods.
However, this requires an independent
estimate for the zero-gravity distance. As we have noted (Chernin et al. \cite{chernin06}),
the size of the group or the zero-velocity distance is a strict lower limit
to $R_{\rm v}$, giving an upper limit to the DE density. But how to obtain an
upper limit to $R_{\rm v}$ and thus the interesting lower limit to DE? In our previous
studies (Chernin et al. \cite{chernin06}, \cite{chernin09}) we have assumed that
the Hubble flow begins beyond $R_{\rm v}$. Here we study
this question more quantitatively
and pay special attention to the distance where the local outflow reaches the global Hubble
rate, as this has particular significance in the point-mass model.

\section{Position of $R_{\rm v}$ in the Hubble diagram}

If the local DE density is equal to the global one, there is the suggestive
coincidence with $R_{\rm v} \approx 1.3 - 1.4$ Mpc being also the distance
where the local Hubble flow begins to be seen.
But one would like to find a way to measure the zero-gravity distance independently,
using local data on the mass and the outflow. What could be its signature in the observations?
In order to clarify this question we have calculated, within the two-component model, the present expected locus
of outflowing dwarf galaxies for several different values of the relevant parameters.

\subsection{The normalized Hubble diagram}

Previously, we introduced the concept
of a normalized Hubble diagram, in order to study the kinematic structure
of a group and its surroundings (Teerikorpi et al. \cite{teerikorpi08}) in the presence of
dark energy. If one fixes the mass of the group and the local dark energy density,
one can calculate the zero-gravity radius $R_{\rm v}$ and the
vacuum Hubble constant $H_{\rm v} = (8\pi G\rho_{\rm v}/3)^{1/2}$. 
Then in the dimensionless representation with
$R/R_{\rm v}$ and $V/H_{\rm v}R_{\rm v}$ as $x$- and $y$-axes, respectively,
one may conveniently describe different dynamical regions of the system (See Fig.\ref{fig1}).

Writing the total mechanical energy as
$E = -\alpha \,GM/R_{\rm v}$,
there is a minimum energy curve corresponding to
$\alpha = \frac{3}{2}$. Test particles ejected from the region of bound orbits
($R < R_{\rm v}$) cannot appear below this curve for $R > R_{\rm v}$.
In fact, the minimum energy curve can be used to give an upper limit to the local
DE density or a lower limit to $R_{\rm v}$ which may be stricter than that obtained from the size
and mass of the system.

The physical sense of this minimum energy curve is that it corresponds to ejections an infinitely
long time ago. Hence, any upper limit for the age provides a still stricter lower-limit
curve. Consider the vacuum Hubble time $T_{\rm v} = 1/H_{\rm v}$, where
\begin{eqnarray}
H_v &=& (8\pi G\rho_{\rm v}/3)^{1/2} \\
    &=& 61.0 \times (\rho_{\rm v}/7\,10^{-30} {\rm g/cm}^3)^{1/2}\,\, {\rm km/s/Mpc}
\end{eqnarray}
The vacuum Hubble time $T_{\rm v}$ is larger than the global Hubble time
by the factor $(1+\rho_m/\rho_{\rm v})^{1/2} = (\Omega_{\rm v})^{-1/2}$ for a flat universe
and for $\rho_{\rm v} =$ the global DE density.
Thus  $T_{\rm v}$ is a natural upper limit
for the age. In the standard model $T_{\rm v} = 16\, 10^9$ yrs and the age of
the universe ($13.6\,10^9$ yrs) is
about $0.85 \times T_{\rm v}$.

\subsection{The present distance-velocity relation}
 
The flight time from the center of the group ($x_0 \ll 1$) to the normalized distance $x = r/R_{\rm v}$,
for a particle with energy $E = -\alpha \, GM/R_{\rm v}$ can be parametrized in terms of
the vacuum Hubble time $T_{\rm v} = 1/H_{\rm v}$:
\be
t = T_{\rm v} \int_{x_0}^x \! (x^2 + 2/x -2\alpha)^{-1/2} \, dx \, ,
\label{integral1}
\ee
while the normalized velocity $y$ is related to $x$ as

\be
y = (x^2 + 2/x - 2\alpha)^{1/2}
\label{y-x}
\ee
One can calculate for each location $x$ the required
energy $\alpha$ for a particle to reach this distance after the flight time $t$,
and thus the normalized velocity $y$.
This is similar to what has been done in Lema\^{i}tre-Tolman
solutions (e.g. Peirani \& de Freitas Pacheco \cite{peirani06}, \cite{peirani08}).

\begin{figure}
\epsfig{file=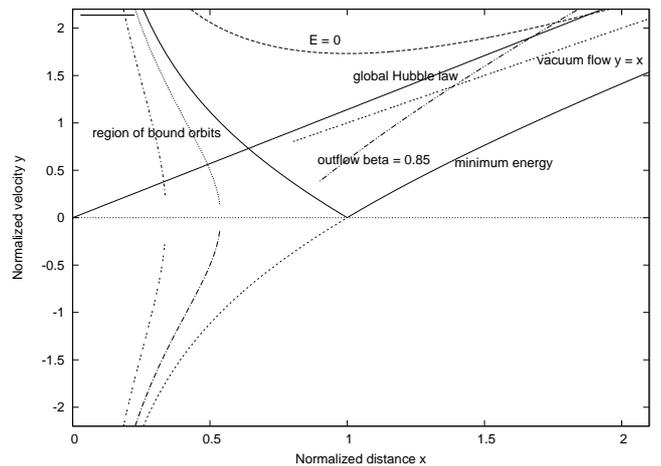, angle=270, width=9.0cm}
\caption{The normalized distance-velocity diagram showing
the predicted local flow for the age of the standard world model, corresponding
to $\beta = t/T_{\rm v} = 0.85$. The local flow reaches the global
Hubble rate around $x \approx 1.7 R_{\rm v}$.
}
\label{fig1}
\end{figure}

As an example, in the normalized diagram of Fig.\ref{fig1} we have indicated the predicted current distance-velocity
curve for the age $t/T_{\rm v} = 0.85$ for the case where the local DE density is equal to the global
one (eq.\ref{integral1} or \ref{integral2} with $\rho_{\rm loc} =\rho_{\rm v}$). The straight line above
the vacuum Hubble flow ($y = x$) is the global Hubble law with $H = (\Omega_{\rm v})^{-1/2} H_{\rm v}$ for
 $\Omega_{\rm v} =
0.77$.

\begin{figure}
\epsfig{file=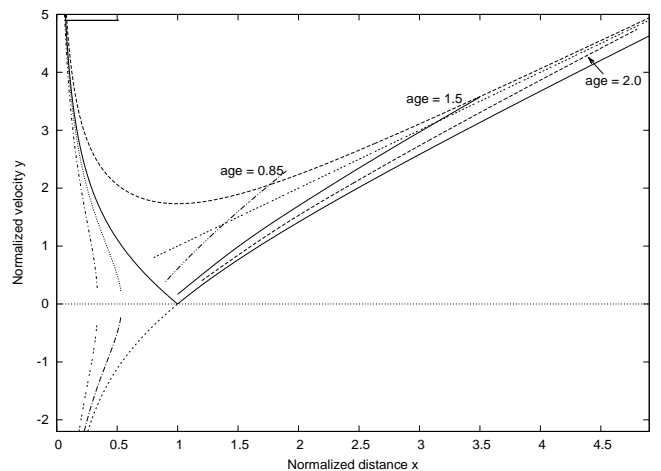, angle=270, width=9cm}
\caption{The normalized distance-velocity diagram showing
the predicted local flow for several flight times: 0.85, 1.5 and 2.0
vacuum Hubble times. For large distances and long times the flow approaches
the vacuum flow $y=x$.
}
\label{fig2}
\end{figure}

In Fig.\ref{fig2} we illustrate how the increasing flight time (0.85, 1.5, and 2.0 $\times T_{\rm v}$)
influences the distance-velocity relation in the normalized Hubble diagram.
Note that at large distances and long
times the flow approches the vacuum flow $y=x$. The vacuum flow is also the asymtotic line
for the zero-energy and the minimum energy curves, as can be seen from eq.\ref{y-x}.

From the  $y = y(x)$ we can obtain
the locus of present distance-velocity positions as ($xR_{\rm v}$, $yH_{\rm v}R_{\rm v}$).
So, fixing the mass $M$ and the DE density $\rho_{\rm v}$ (which give the  distance $R_{\rm v}$ and
 the vacuum Hubble
constant $H_{\rm v}$ needed in the normalization), and the flight time $t/T_{\rm v}$ (say, the age of the universe)
we can calculate the velocity-distance relation for the ordinary Hubble diagram.
For example, in Chernin et al. (\cite{chernin09}) the mass was varied and the predicted Hubble relation was
compared with the outflow around the Local Group, assuming that the local DE density is the same as the global one.

\subsection{Times spent in gravitation and DE dominated regions}

We see from Fig.\ref{fig2} that in the standard model the present Hubble relation deviates from the vacuum flow
$y =x$ and, as mentioned, only in the distant future it approaches
that relation for the DE dominated cosmos. In fact, the galaxies presently located in the range $1 \la
x \la 1.7$ have spent in the past most of their time within the gravity dominated region and less than half
in the DE dominated zone. For the standard model (age = 0.85 vacuum Hubble time),
it is only at $x \ga 1.8$ where the galaxies have spent about equal or longer time in the DE dominated
region. This is why the current Hubble relation has not had time to approach closer to the vacuum
flow, even though the static zero-energy border has existed since the formation of the group.

\subsection{When the local DE density differs from the global value}

We see from Fig.\ref{fig1} that the Hubble ratio of the local flow aproaches the global value around $R \approx 1.7 R_{\rm v}$.
This suggests the location where one can bracket from upwards the possible values of $R_{\rm v}$. But is
this criterion -- where the observed local flow is close to the global rate -- valid also for
dark energy densities smaller than the global one?
 
\begin{figure}
\epsfig{file=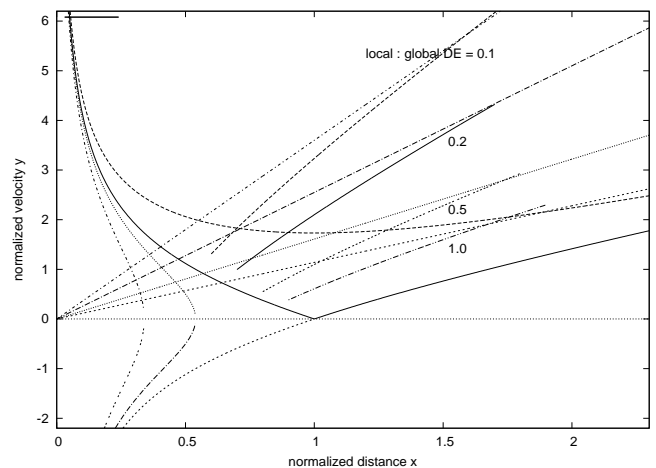, angle=270, width=9cm}
\caption{The normalized distance-velocity diagram
for different values of the local-to-global
dark energy density ratio. Note that the predicted
local flow reaches the global Hubble rate (the straight
lines beginning from the origo) around 1.5 -- 1.7 times the zero-gravity distance.}
\label{fig3}
\end{figure}

In Eq.\ref{integral1} the time $T_{\rm v}$ actually means the vacuum Hubble time corresponding to the local
DE density $\rho_{\rm loc}$. As we want to use the global vacuum Hubble time (related to the standard model)
as a convenient unit also when the local and global densities differ, we write:

\be
t/T_{\rm v} = (\rho_{\rm v}/\rho_{\rm loc})^{1/2} \int_{x_0}^x \! (x^2 + 2/x -2\alpha)^{-1/2} \, dx \, ,
\label{integral2}
\ee
In the normalized diagram of  Fig.\ref{fig3} we show the calculated local Hubble relations for several cases
of the local-to-global DE ratios,
$\rho_{\rm loc}/\rho_{\rm v} =$ 1, 0.5, 0.2, and 0.1. For each curve we show the corresponding global
Hubble line
(the same in ordinary units, but differing in the normalized presentation).

It is seen that in each case the local flow approaches the global Hubble law well beyond the zero-gravity
radius $R_{\rm v}$, in fact at about the same normalized $x$-location. In the ordinary Hubble diagram
these locations may differ much (due to the different values of $R_{\rm v}$), but it is important
that the criterion which we have used in our previous papers (e.g. Chernin et al. \cite{chernin06},
\cite{chernin09}) thus obtains quantitative support: the known global Hubble law allows
one to define a robust upper limit to $R_{\rm v}$ from the local expansion flow.

\section{Concluding remarks}

It is interesting to realize that the mass concentrations have had around them the zero-gravity
surface since the formation of the concentration.\footnote{The total mass of the system within $R_{\rm V}$
may vary
with time after its formation because of accretion of material from outside or escape of dwarf galaxies
from the system. In reality such mass variations can hardly be very significant. In any case, the value
of the zero-gravity radius should follow the total mass in accordance with Eq.1.}

Beyond this surface the acceleration is positive,
while within it there is deceleration. We cannot directly measure the accelerations in order to define
the location of the zero-gravity distance $R_{\rm v}$ and hence determine the quantity $\rho_{\rm loc}/M$ (and
hence the dark energy density, if the mass $M$ is known). However, as we have previously argued
and in more detail shown here,
we can obtain information about $R_{\rm v}$ by comparing the current behaviour of the local
flow with the global Hubble law. The latter is related to the global time scale and is known from
various large-scale determinations.

\subsection{The local outflow and the zero-acceleration redshift $z_{\rm v}$}

The calculations show that the local flow should reach the global expansion rate around
a distance 1.5 -- 1.7 times the zero-gravity radius (for the local DE density from 0.1 to 1 global
density). This allows one to set an upper limit $R_2$ to $R_{\rm v}$ and hence a lower limit to
$\rho_{\rm loc}/M$. If the mass-point model is inadequate, so that in addition to
the known mass $M$ of the group itself there is some unknown extended mass
$\Delta M$
up to $R_{\rm v}$, then the lower limit to  $\rho_{\rm loc}$ from eq.2 is just made more robust:
$\rho_{\rm loc}/\rho_{\rm v}> [(M+\Delta M)]/0.9R_2^3 > M/0.9R_2^3$ (in units of $10^{12} \Msun$ and Mpc).

What happens at larger distances depends on whether the local model is still valid -- 
one finally reaches 
the distance $R_{\rm M}$ which gives the size of the volume from which
the mass $M$ has been gathered during the formation of the group (so-called Einstein-Straus vacuole;
Chernin et al. \cite{chernin06}).
One can calculate the present radius $R_{\rm M}$ of the region from which the mass $M$ was gathered,
assuming the present average cosmic mass density $\rho_m$. 
It is 
$R_{\rm M} = (2 \rho_{\rm loc}/\rho_m)^{1/3} R_{\rm v}$
(e.g. $= 1.7 R_{\rm v}$ for $\Omega_{\rm v}/\Omega_{\rm m} = 0.77/0.23$).

Interestingly, this is the same 1.7 as in the global scale factor ratio leading to $z_{\rm v} = 0.7$.
Namely, the requirement that the global acceleration is zero when $\rho_{\rm v} = 0.5 \rho_{\rm m}(z_{\rm v})$
leads
in terms of the current mass density to the condition $2\rho_{\rm v} = (1+z_{\rm v})^3 \rho_{\rm m}$ or
$(2\rho_{\rm loc}/\rho_{\rm m})^{1/3} = (1+z_{\rm v}) =1.7$. 
This is also the distance where the global Hubble ratio is reached, because
at this point the enclosed mass is the same as for the uniform global Friedmann model, hence
the expansion rate is the same.

To repeat, when
the local dark energy density is equal to the global DE density, the
global Hubble rate is reached at the distance $(1+z_{\rm v})R_{\rm v}$. This distance is also
the same as the radius of the ideal Einstein-Straus vacuole $R_{\rm M}$.

\subsection{An application to the Local Group}

An implication is that the best place for the study of the local dark energy using
the present approach would be where the point mass plus void model extends beyond
the standard radius $R_{\rm M} = (2 \rho_{\rm loc}/\rho_{\rm m})^{1/3} R_{\rm v}$.
Interestingly, this may just be the case in the Local Volume where, 
according to Karachentsev et al. (\cite{karachentsev09}), there seems to
be an average underdensity by a factor of 2 or 3 locally, 
and the vacuole from which the central mass was gathered, may then exceed $R_{\rm M}$
by a factor of 1.26 or 1.44, respectively.
 
If the local flow reaches the global Hubble rate
around the distance $R_2$, as defined by sufficiently accurate data, and if $R_2 \la R_{\rm M}$, then robustly

\be
\frac{R_2}{2} < R_{\rm v} < R_2
\ee
For example, if we take for the Local Group $R_2 \approx 2.4$ Mpc
and $M = 2\times 10^{12} M_{\odot}$ (Karachentsev et al. \cite{karachentsev09}), then we obtain using eq.2 

\be 0.5 < \rho_{\rm loc}/\rho_{\rm v}  < 2.0. \ee
Expressed in another way, for $M = 2.0 - 3.5\times 10^{12} M_{\odot}$ (see Chernin et al. \cite{chernin09}) 
and $\rho_{\rm loc} = \rho_{\rm v}$,
the zero-gravity radius is 1.3 to 1.55 Mpc and the predicted distance where
the local flow reaches the global Hubble rate is 2.2 -- 2.6 Mpc, which
agrees with the observations (Karachentsev et al. \cite{karachentsev09}).
Hence, as in our earlier
estimations, the value of the local density proves to be near the
global one, supporting
the universal nature of the dark energy having the same density
both globally and locally.

\acknowledgements
 A.C. thanks the RFBR for partial support via the grant 10-02-00178.
We also thank the anonymous
referee for useful comments.

\end{document}